\documentclass[11pt,a4paper]{article}
\usepackage[a4paper,left=3cm,right=3cm]{geometry}
\usepackage{graphicx}
\usepackage{amsmath}
\usepackage{authblk}
\usepackage{hyperref}
\usepackage{caption}
\usepackage{subcaption}
\usepackage{textcomp}   

\usepackage{lineno}

\title{\bf Classifying hadronic objects in ATLAS with ML/AI algorithms}

\author{
    {\large Leonardo Toffolin$^{1,2,3}$ on behalf of the ATLAS Collaboration \thanks{\copyright ~Copyright 2025 CERN for the benefit of the ATLAS Collaboration. CC-BY-4.0 license.}}\\
    {\normalsize
    $^{1}$ University of Trieste, Italy\\
    $^{2}$ INFN Trieste, Gruppo Collegato di Udine, Italy\\
    $^{3}$ European Organisation for Nuclear Research (CERN), Geneva, Switzerland}
}

\date{Proceedings of the 32nd International Symposium on Lepton Photon\\
Interactions at High Energies (LP2025),\\
25–29 August 2025, Madison, WI, USA}

\begin{document}
\maketitle

\begin{abstract}
    The identification of hadronic final states plays a crucial role in the physics programme of the ATLAS Experiment at the CERN LHC. Sophisticated artificial intelligence (AI) algorithms are employed to classify jets according to their origin, distinguishing between quark- and gluon-initiated jets, and identifying hadronically decaying heavy objects such as $W$ bosons and top quarks. This contribution summarises recent developments in constituent-based tagging architectures, including graph neural networks (GNNs) and transformer-based approaches, their performance in simulated and real data, and future perspectives towards data-driven optimisation and model-independent tagging strategies.
\end{abstract}

\section{Machine Learning for jet tagging}

Hadronic jets are among the most abundant final-state objects in proton–proton collisions at the CERN LHC. Classifying them according to their initiating particle is of high importance for precision measurements, searches for new phenomena, and QCD studies. 

The ATLAS Collaboration~\cite{ATLAS} has developed dedicated taggers to distinguish between quark- and gluon-initiated jets, identify heavy-flavour jets, and recognise boosted hadronic decays of $W$ bosons and top quarks. Recent advances in machine learning (ML) have shifted jet-tagging approaches from high-level substructure observables to constituent-based architectures that directly use the four-vectors of jet constituents. These include:
\begin{itemize}
    \item Fully-connected deep neural networks (FC DNNs): a straightforward approach using ordered constituent features.
    \item Energy and Particle Flow Networks ({\tt EFN}, {\tt PFN})~\cite{EFN PFN}: point-cloud models, based on the DeepSets approach, enforcing permutation invariance. Unlike {\tt PFN}, the {\tt EFN} model uses only infrared and collinear-safe variables.
    \item Graph Neural Networks (GNNs): models, such as {\tt ParticleNet}~\cite{ParticleNet}, where jet constituents are represented as nodes connected by learned geometric relations.
    \item Transformers: architectures inspired by natural language processing, using attention mechanisms to dynamically relate jet constituents.
\end{itemize}

These approaches exploit the full detector granularity and enable data-driven learning of correlations that characterise different jet types, outperforming traditional observables-based taggers.

\section{Quark/gluon and flavour tagging}

\begin{figure}
    \centering
    \includegraphics[width=0.58\linewidth]{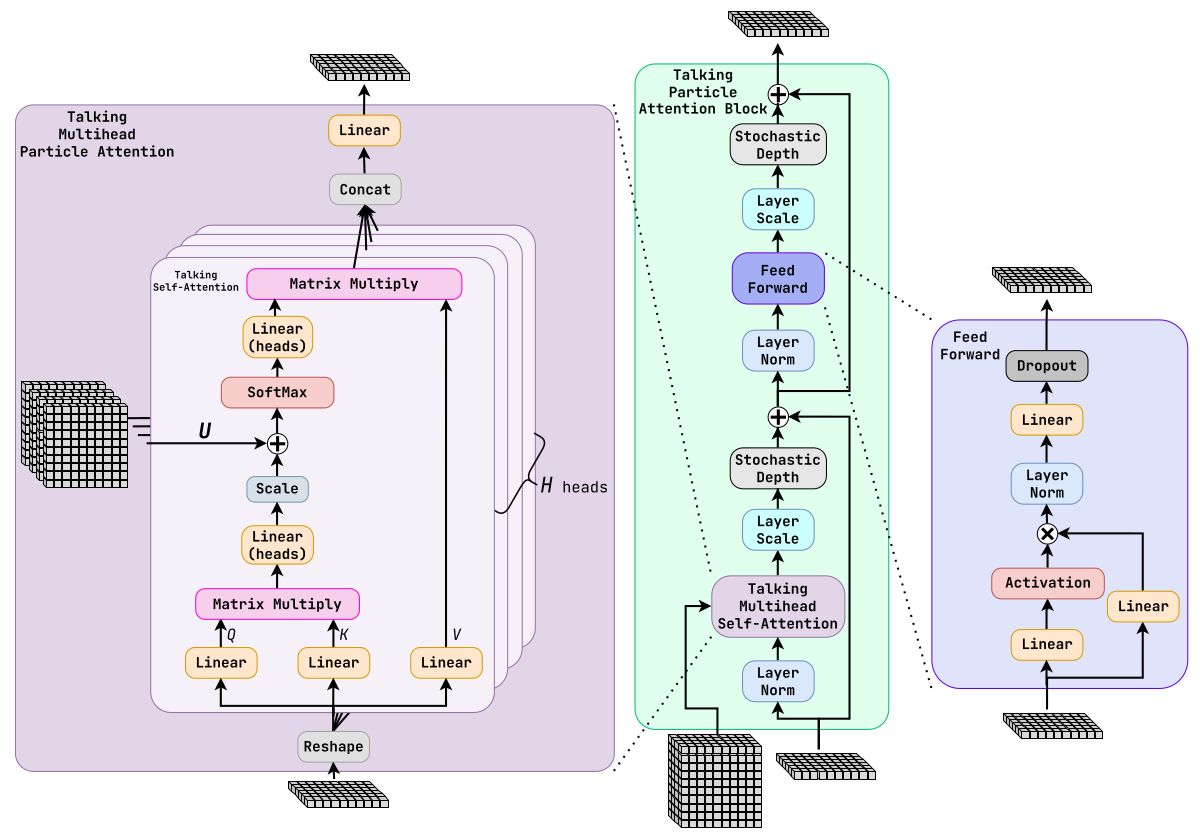}
    \caption{Architecture of the {\tt DeParT} algorithm~\cite{DeParT}.}
    \label{fig:DeParT architecture}
\end{figure}

\begin{figure}[h!]
    \centering
    \begin{subfigure}[c]{0.493\textwidth}
    \includegraphics[width=\textwidth]{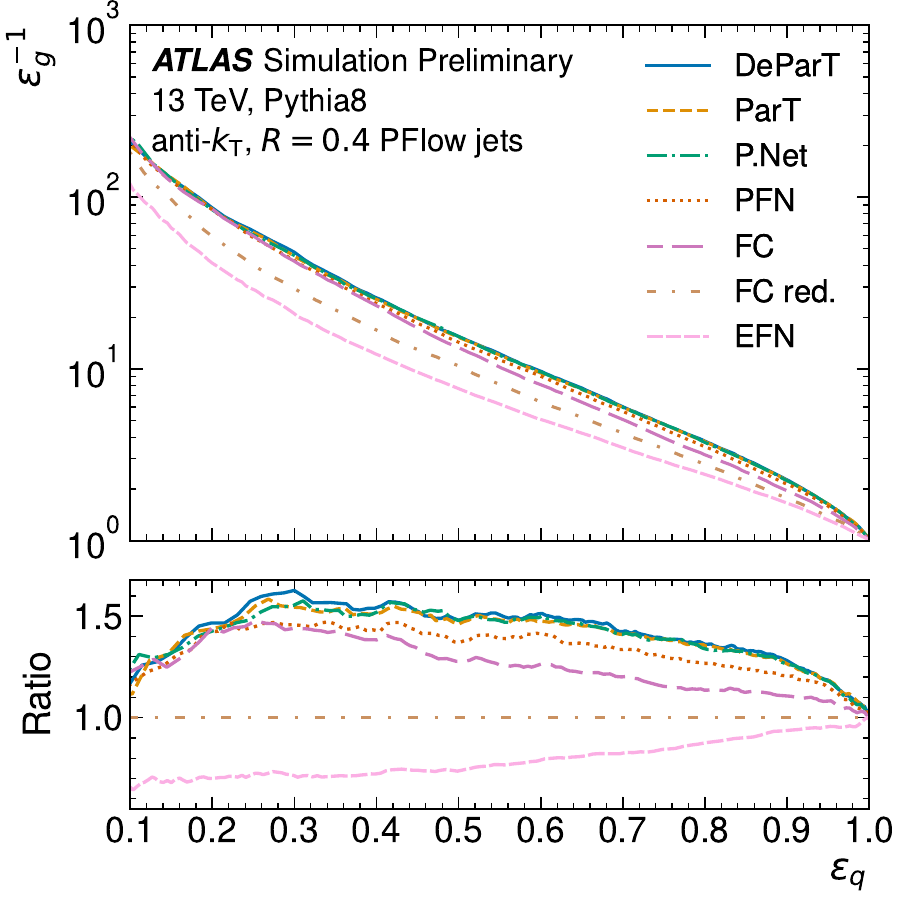}
    \caption{}
    \end{subfigure}
    \hfill
    \begin{subfigure}[c]{0.499\textwidth}
    \includegraphics[width=\textwidth]{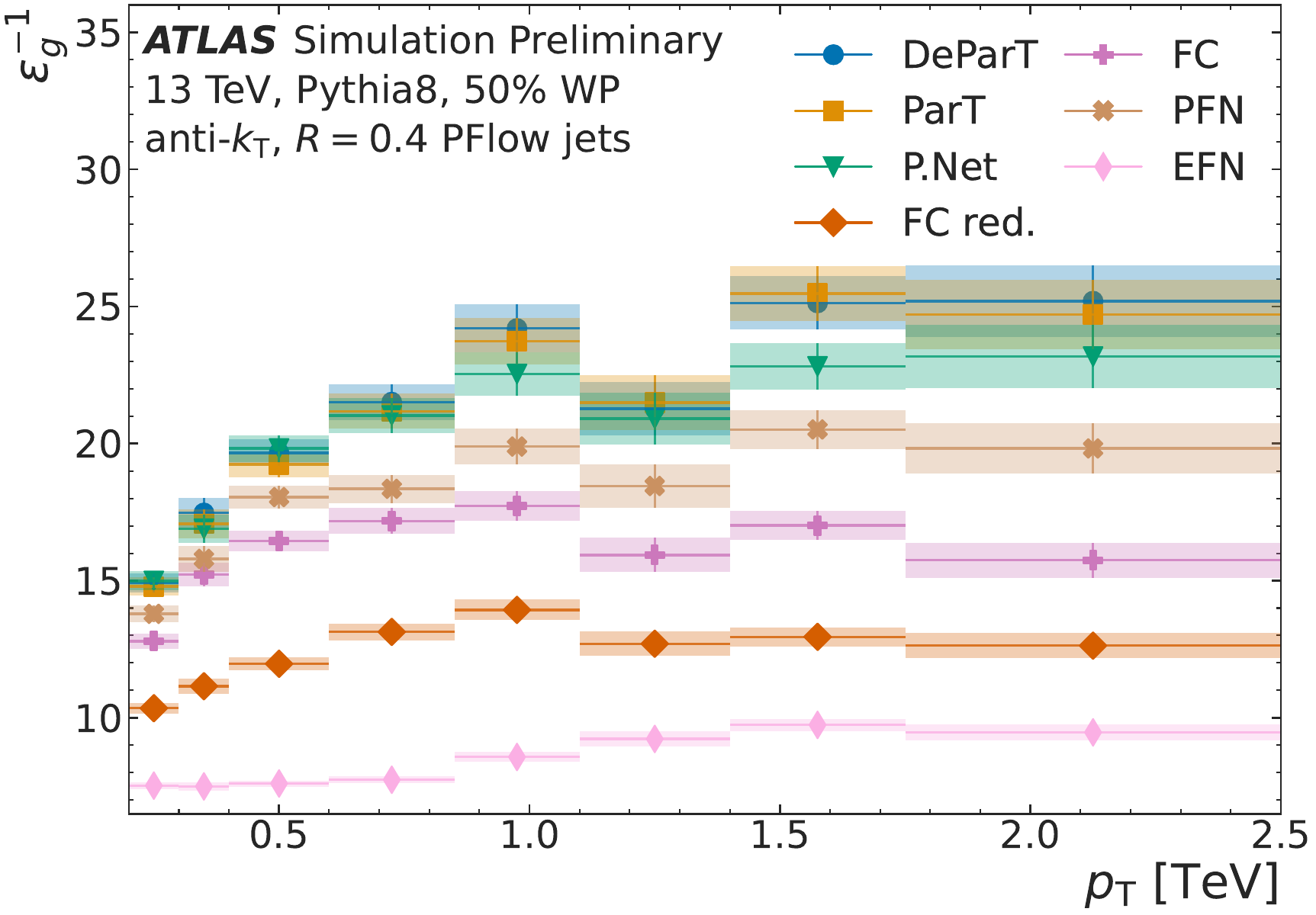}
    \caption{}
    \end{subfigure}
    \caption{Performance of the {\tt DeParT} transformer-based algorithm for quark/gluon tagging compared with FC DNN, {\tt PFN}, {\tt EFN}, and {\tt ParticleNet} architectures, expressed in terms of gluon-jet rejection $\epsilon^{-1}_{g}$. The rejection is shown as a function of (a) quark identification efficiency $\epsilon_{q}$ and (b) $p_{\rm T}$. Taken from~\cite{DeParT}.}
    \label{fig:DeParT}
\end{figure}

Identification and discrimination of quark/gluon ($q$/$g$) jets are among the most challenging classification tasks, due in part to the strong dependence on non-perturbative effects and parton-shower modelling. Those initiated by gluons have on average more constituents and a broader collimation pattern than quark-initiated ones. In the ATLAS Collaboration, $q$/$g$ tagging has historically relied on the number of tracks and track width, combined in a boosted decision tree (BDT). This approach has gradually been replaced by more modern ML architectures, based on GNNs or transformers~\cite{Particle transformer}. In particular, the \emph{Dynamically Enhanced Particle Transformer} algorithm, termed {\tt DeParT}~\cite{DeParT}, represents the current state of the art for the identification of the small-radius ($R=0.4$) jets. It operates on \emph{particle flow objects} (PFOs) and uses both kinematic and relational features among constituents, with attention blocks to replace the BDT. As shown in Figure~\ref{fig:DeParT}, {\tt DeParT} achieves improved gluon-jet rejection across a broad $p_{\mathrm{T}}$ range compared to fully connected (FC) networks or energy flow networks~\cite{DeParT}.

The success of transformers architecture for jet classification is also testified by the adoption of the {\tt GN2} algorithm~\cite{GN2} for jet flavour tagging in the ATLAS Collaboration. This algorithm exploits track-level information, and auxiliary tasks such as vertex and track-origin classification are employed to stabilise the training. {\tt GN2} demonstrates a factor of about three improvement in light-jet rejection compared to the Run-2 ATLAS baseline taggers {\tt DL1r} and {\tt DL1d}~\cite{FTAG algos Run-2}.

\section{$W$-boson and top-quark tagging}

\begin{figure}
    \centering
    \begin{subfigure}[c]{0.499\textwidth}
    \includegraphics[width=\textwidth]{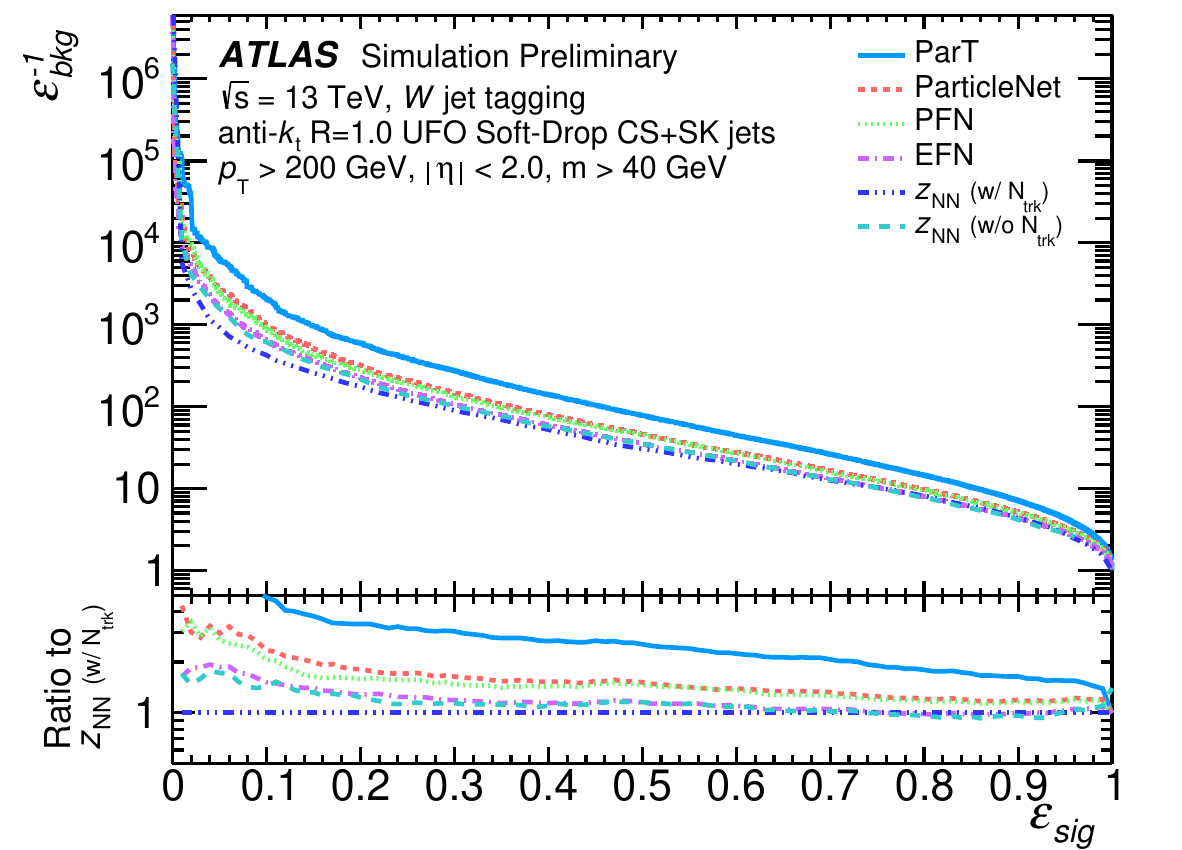}
    \caption{}
    \end{subfigure}
    \hfill
    \begin{subfigure}[c]{0.493\textwidth}
    \includegraphics[width=\textwidth]{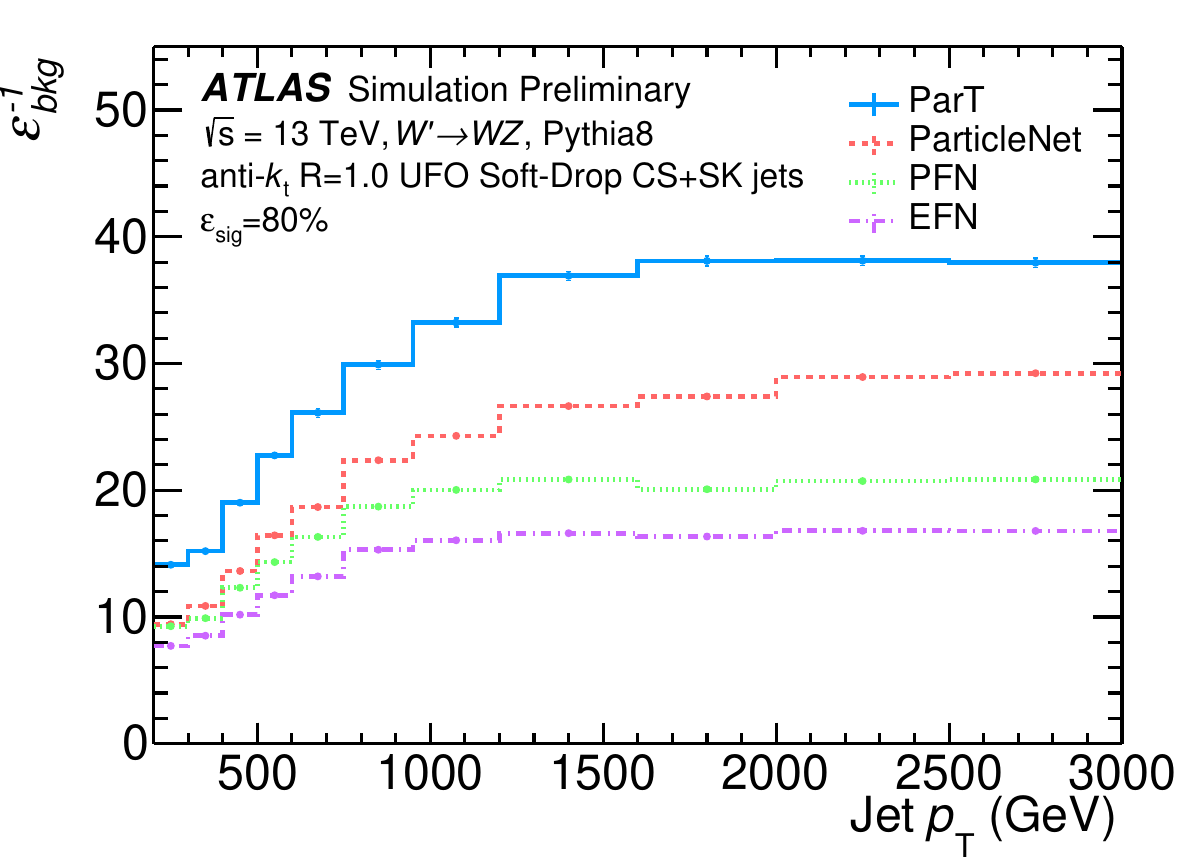}
    \caption{}
    \end{subfigure}
    \caption{Performance of the {\tt ParT} algorithm as a function of the jet identification efficiency $\epsilon_{\rm sig}$ (a) and jet $p_{\rm T}$ (b), compared with {\tt EFN}, {\tt PFN} and {\tt ParticleNet}. Taken from~\cite{W tagging with constituents}.}
    \label{fig:W tagging constituents}
\end{figure}

At higher transverse momenta, hadronically decaying $W$ bosons and top quarks are reconstructed as single large-radius ($R=1.0$) jets. The constituent-based particle transformer {\tt ParT}, trained and evaluated on $W' \to WZ$ and multi-jet events, is used as the state-of-art algorithm for $W$-jet tagging~\cite{W tagging with constituents}, and its performance can be compared with the established {\tt EFN}, {\tt PFN} and {\tt ParticleNet} taggers (Figure~\ref{fig:W tagging constituents}).

\begin{figure}[h!]
    \centering
    \begin{subfigure}[c]{0.498\textwidth}
    \includegraphics[width=\textwidth]{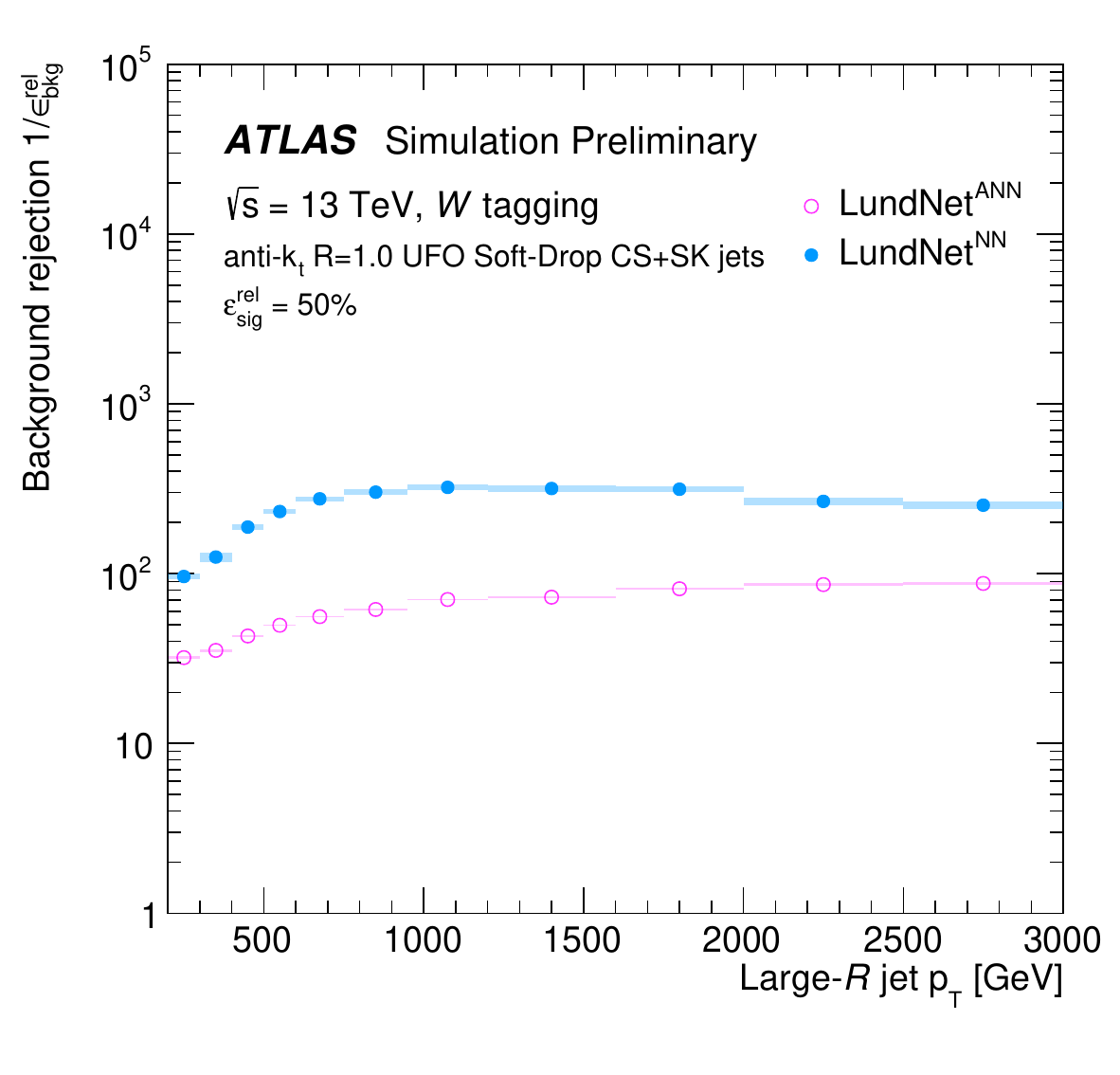}
    \caption{}
    \end{subfigure}
    \hfill
    \begin{subfigure}[c]{0.493\textwidth}
    \includegraphics[width=\textwidth]{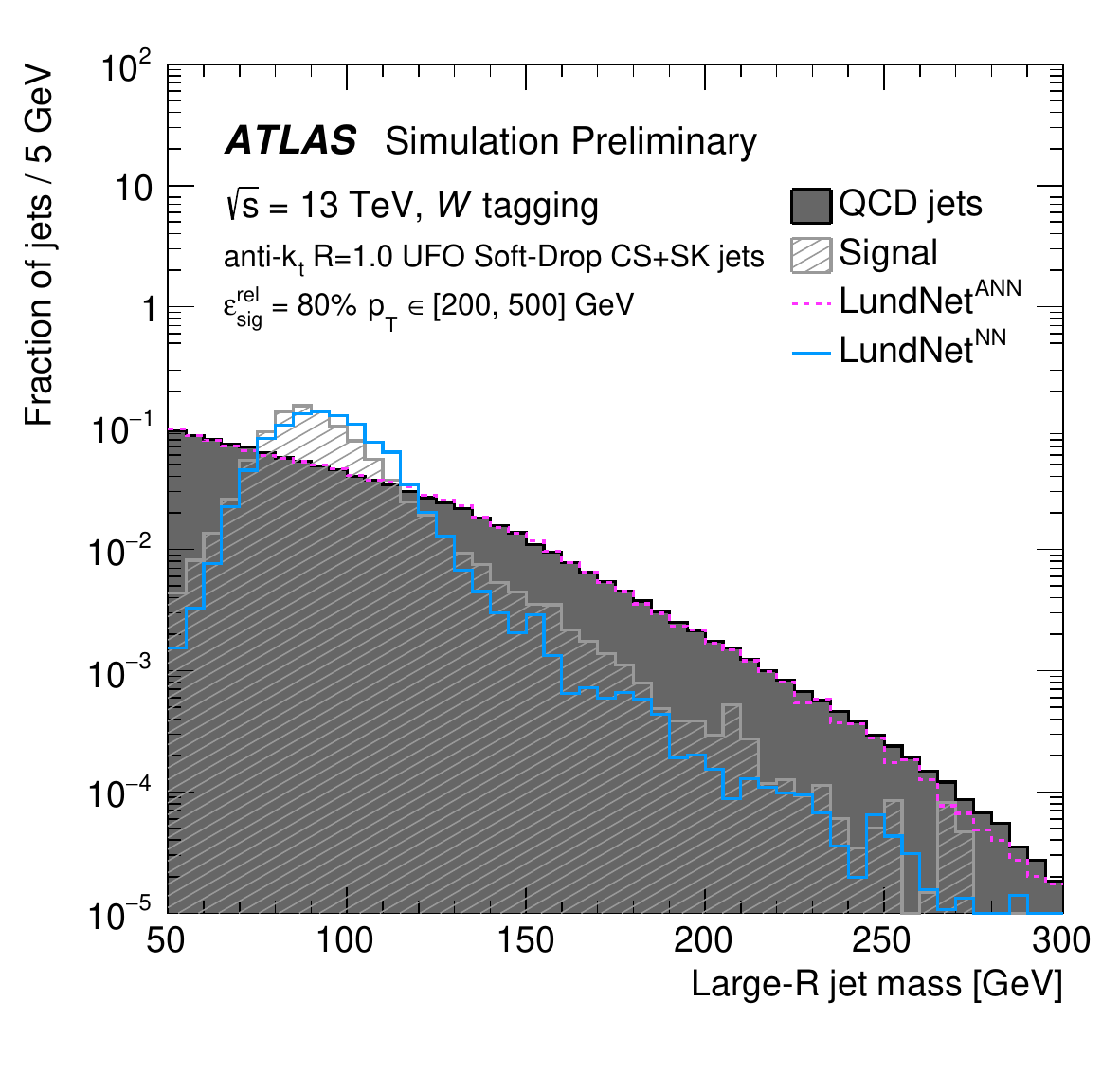}
    \caption{}
    \end{subfigure}
    \caption{(a) Background rejection performance of the {\tt LundNet} and {\tt LundNetANN} taggers for $W$-boson identification. The {\tt LundNetANN} algorithm, despite a smaller rejection power, has the ability of decorrelating the performance from jet mass and reproducing better the QCD jets contribution, as shown in (b). Taken from~\cite{W tagging LJP}.}
    \label{fig:W tagging LJP performance}
\end{figure}

Constituent-based models such as {\tt LundNet}~\cite{W tagging LJP}, a GNN exploiting the jet clustering history in the Lund Jet Plane (LJP), have achieved state-of-the-art performance in this regime. Variants like {\tt LundNetANN} incorporate adversarial training to decorrelate the tagger output from jet mass, reducing systematic uncertainties in background modelling (Figure~\ref{fig:W tagging LJP performance}). Nonetheless, model dependence with respect to MC generators remains significant: performance reductions up to 40\% are observed when varying parton-shower models.

In addition to $W$-boson identification, top-quark tagging exploits the characteristic three-prong substructure of hadronic top decays reconstructed as large-radius ($R=1.0$) jets. The constituent-based {\tt ParticleNet} algorithm has achieved state-of-the-art performance in this regime, providing superior accuracy across a wide $p_{\mathrm{T}}$ range albeit with higher computational complexity~\cite{Boosted top tagging}. Alternative GNN approaches, such as {\tt LundNet}, further improve robustness against variations in Monte Carlo generators.

\section{Summary and Outlook}

The ATLAS Collaboration continues to develop advanced ML approaches for the classification of hadronic objects. Constituent-based methods, including GNNs and transformers, have become the leading architectures for modern jet tagging, achieving unprecedented accuracy while revealing challenges related to generator dependence and systematic robustness. Future work focuses on data-driven performance validation, uncertainty mitigation, and hybrid models combining complementary representations, such as the constituent-based transformers and clustering-tree-based GNNs. 


\begin{thebibliography}{9}

\bibitem{ATLAS}
ATLAS Collaboration, \emph{The ATLAS Experiment at the CERN Large Hadron Collider}, \href{https://doi.org/10.1088/1748-0221/3/08/S08003}{JINST \textbf{3} (2008) S08003}.
\bibitem{EFN PFN}
P.~T.~Komiske, E.~M.~Metodiev and J.~Thaler, \emph{Energy Flow Networks: Deep Sets for Particle Jets}, \href{https://doi.org/10.1007/JHEP01(2019)121}{JHEP {\bf 01} (2019) 121}.
\bibitem{ParticleNet}
H.~Qu and L.~Gouskos, \emph{ParticleNet: Jet Tagging via Particle Clouds}, \href{https://doi.org/10.1103/PhysRevD.101.056019}{Phys. Rev. D {\bf 101} (2020) 056019}.
\bibitem{Particle transformer}
H.~Qu, C.~Li, S.~Qian, \emph{Particle Transformer for Jet Tagging}, arXiv: \href{https://doi.org/10.48550/arXiv.2202.03772}{2202.03772}.
\bibitem{DeParT}
ATLAS Collaboration, \emph{Constituent-Based Quark Gluon Tagging using Transformers with the ATLAS detector}, ATL-PHYS-PUB-2023-032, {\sc url}: \url{https://cds.cern.ch/record/2878932/}.
\bibitem{GN2}
ATLAS Collaboration, \emph{Transforming jet flavour tagging at ATLAS}, arXiv: \href{https://doi.org/10.48550/arXiv.2505.19689}{2505.19689}, 2025.
\bibitem{FTAG algos Run-2}
ATLAS Collaboration, \emph{ATLAS flavour-tagging algorithms for the LHC Run 2 pp collision dataset}, \href{https://doi.org/10.1140/epjc/s10052-023-11699-1}{Eur. Phys. J. C {\bf 83} (2023) 7, 681}.
\bibitem{W tagging with constituents}
ATLAS Collaboration, \emph{Constituent-Based $W$-boson Tagging with the ATLAS Detector}, ATL-PHYS-PUB-2023-020, {\sc url}: \url{https://cds.cern.ch/record/2866592/}.
\bibitem{W tagging LJP}
ATLAS Collaboration, \emph{Tagging boosted $W$ bosons applying machine learning to the Lund Jet Plane}, ATL-PHYS-PUB-2023-017, {\sc url}: \url{https://cds.cern.ch/record/2864131/}.
\bibitem{Boosted top tagging}
ATLAS Collaboration, \emph{Accuracy versus precision in boosted top tagging with the ATLAS detector}, \href{https://doi.org/10.1088/1748-0221/19/08/P08018}{JINST {\bf 19} (2024) P08018}.

\end{thebibliography}
\end{document}